\documentclass[review]{elsarticle}

\usepackage{setspace}
\usepackage{algorithm}
\usepackage{algpseudocode}
\usepackage{multirow}
\usepackage{color}
\usepackage{paralist}
\usepackage{subfigure}
\usepackage{graphicx}
\usepackage{amsmath}
\usepackage{balance}
\usepackage{tabularx}
\usepackage{soul}
\usepackage{microtype}
\usepackage[flushleft]{threeparttable}
\usepackage{lipsum}                     
\usepackage{xargs}                      
\usepackage[colorinlistoftodos,prependcaption,textsize=tiny]{todonotes}
\usepackage{amsfonts}
\usepackage{rotating}
\usepackage{booktabs}

\usepackage{url}
\makeatletter
\g@addto@macro{\UrlBreaks}{\UrlOrds}
\makeatother

\makeatletter
\def\ps@pprintTitle{%
 \let\@oddhead\@empty
 \let\@evenhead\@empty
 \def\@oddfoot{This is authors accepted copy, for final version please refer to DOI: 10.1016/B978-0-12-805303-4.00012-5}%
 \let\@evenfoot\@oddfoot}
\makeatother

\begin{document}
\newcommand*\pct{\scalebox{.9}{\%~}}

\begin{frontmatter}

\title{Investigating Storage as a Service Cloud Platform: \\ pCloud as a Case Study}

\author{Tooska Dargahi$^{\rm a}$
\vspace{6pt}, Ali Dehghantanha$^{\rm b}$ 
\vspace{6pt}, and Mauro Conti$^{\rm c}$
\\\vspace{6pt}  $^{a}${\em{Department of Computer Engineering, West Tehran Branch, Islamic Azad University, Iran}};\\
$^{b}${\em{The School of Computing, Science \& Engineering, University of Salford, United Kingdom}};\\
$^{c}${\em{Department of Mathematics, University of Padua, Italy}}}

\begin{abstract}
Due to the flexibility, affordability and portability of cloud storage, individuals and companies envisage the cloud storage as one of the preferred storage media nowadays. 
This attracts the eyes of cyber criminals, since much valuable information such as user credentials, and private customer records are stored in the cloud. There are many ways for criminals to compromise cloud services; ranging from non-technical attack methods, such as social engineering, to deploying advanced malwares. Therefore, it is vital for cyber forensics examiners to be equipped and informed about best methods for investigation of different cloud platforms. In this chapter, using \textit{pCloud} (an extensively used online cloud storage service) as a case study, and we elaborate on different kinds of artefacts retrievable during a forensics examination. We carried out our experiments on four different virtual machines running four popular operating systems: a 64 bit \textit{Windows~8}, \textit{Ubuntu~14.04.1 LTS}, \textit{Android~4.4.2}, and \textit{iOS~8.1}. Moreover, we examined cloud remnants of two different web browsers: \textit{Internet Explorer} and \textit{Google Chrome} on Windows. We believe that our study would promote awareness among digital forensic examiners on how to conduct cloud storage forensics examination.
\end{abstract}


\end{frontmatter}


\section{Introduction} \label{sec:intro}

The usage of cloud storage, among individuals and companies, is increasing day by day. Due to the recent report of the Forbes (2015), \textit{``47\pct of marketing departments will have at least 60\pct of their applications on a cloud platform by 2017"}~\cite{Columbus2015predicting}, and  
\textit{``cloud market cap will pass \$500 billion by 2020"}~\cite{Konard2015report}. Even though cloud storage offers several advantages compared to traditional and local storage of data, cloud users are concerned about the integrity of stored data, security and user privacy issues~\cite{chen2012data, ardagna2014anonymous}. There exist several solutions which could be considered by security experts in order to protect the stored data, and preserve privacy of the cloud users \cite{takabi2010security, mathew2012survey, hosseinzadeh2015security, memarian2015eyecloud}. Adopting security mechanisms is useful in protecting data against being modified and accessed by unauthorized users, and make it difficult for the attackers to abuse the data. However, the artefacts which potentially remain on the cloud storage servers could threaten privacy of the cloud users. In such a case, security mechanisms might not suffice to preserve users' privacy. As a result, protecting the sensitive data against cloud storage services, which leak the privacy of the users, is trending as an issue to the law enforcement agencies and other digital forensic investigators. Moreover, it should be contemplated that organized cyber criminals are always able to find new ways of evading the rules~\cite{choo2008criminal, choo2008organised}. 
This motivated several researchers to conduct a number of cloud storage forensic investigations on various cloud services and applications~(apps)~\cite{quick2013cloud, quick2013forensic}. 
However, with the ever increasing introduction of such cloud services and technologies, 
having an up to date understanding of possible data remnants after using new cloud storage applications is fundamental for forensic practitioner~\cite{daryabar2013review}.


In this chapter, we consider \textit{pCloud}\footnote{\url{https://www.pcloud.com/}} as a case study to identify the possible evidential data that may remain after the use of pCloud on several different computer systems. 
pCloud is a free online cloud storage service (founded in 2013 in Switzerland), which has over four million users right today~\cite{pcloud2016user}.  pCloud users are able to store, sync and share their files, as well as make backup from other cloud services such as Dropbox. pCloud provides client-side encryption such that the data, which are leaving the client's system, are encrypted.
Moreover, pCloud has the Quality Management Systems (ISO 9001:2008) and Information Security Management Systems (ISO 27001:2013) certificates.
Due to the increasing use of the pCloud, and several good reviews that it received from the cloud expert reviewers~\cite{CloudswaveAwards, findmysoft, pCloudreview}, we are focusing on probable privacy issues of pCloud in this chapter. To the best of our knowledge, this is the first forensics investigative study of pCloud. In particular, We will answer the following questions in the rest of the chapter:

\begin{itemize}
\item What data (and the location of the data) can be found on \textit{Windows}, \textit{Ubuntu}, \textit{Android}, and \textit{iOS} operating systems when using pCloud services?
\item What data can be leaked while accessing the pCloud using \textit{Google Chrome} and \textit{Internet Explorer} browsers on Windows operating systems?
\item What data of forensic interest can be discovered in live memory on the aforementioned platforms?
\item What data can be captured from network traffic?
\end{itemize}

Before introducing our research methodology and contribution of the chapter, we provide a brief literature review on forensic investigation of cloud storage services.
 
\subsection{Related Work} \label{subsec:related}
Computer system users produce a great deal of digital data day by day in such a way that by 2020, the amount of produced data will exceed 40 zettabytes~\cite{Lucas2012digitaldata}. Therefore, in order to store such a data on cloud, we need to have more fast and secure synchronization between servers and PCs; for which services such as \textit{BitTorrent} are very common these days. In~\cite{farina2014bittorrent}, Farina et al. conducted a forensic investigation on the BTSync client application, and recognized the digital artefacts and network findings which could be then used by digital forensic examiners as an evidence.
Due to the increasing use of cloud computing and cloud storage services, researchers believe that cloud computing is more vulnerable to security and privacy issues, such as information theft~\cite{choo2010cloud, galante2011sony, symantec2011trojan, Duke2014}, in particular considering online cloud services~\cite{taylor2011forensic}. Thus, there is a surge of interest by forensic professionals and privacy experts in cloud forensic analysis in recent years. In this section, we briefly review the state-of-the-art in digital forensics investigation of cloud privacy. 

Compared to the other aspects of computer analysis, only a few research studies have been conducted on cloud storage privacy investigation.  Martini and Choo~\cite{martini2013cloud} were the first to carry out the cloud forensics investigation. They analyzed the \textit{ownCloud} as a case study, in order to find client and server side artefacts that could be useful as evidential data for forensics practitioners in performing cloud analysis. 
With the gradual increase of Cloud storage services, there is a growing tendency among individuals and organizations in using such a service in order to store and access several different kinds of data. Therefore, most of the investigations on cloud context are concentrated on analyzing the privacy leakage probability of the widely used cloud storage services.  For example, Quick and Choo analyzed the process of gathering data, browsing of data and synchronization of files focusing on \textit{Dropbox}~\cite{quick2013dropbox}, \textit{Microsoft SkyDrive}~\cite{quick2013digital}, and \textit{Google Drive}~\cite{quick2014google}. In~\cite{quick2013digital}, the authors found the terrestrial artefacts which are left behind when using SkyDrive on different devices such as mobile phones and desktop computers. Similarly, Quick and Choo studied the possible data remnants on a Windows~7 computer and an Apple iPhone 3G when a user adopts \textit{Dropbox}~\cite{quick2013dropbox} or \textit{Google Drive}~\cite{quick2014google} in order to store, upload, and access data in the cloud.

           

Along the same line of study, Hale~\cite{hale2013amazon} analyzed the digital artefacts remnant on a computer after accessing or manipulating \textit{Amazon} Cloud Drive. They could recover several information, such as installation path, and upload/download operations. 
In~\cite{chung2012digital}, Chung et al. presented new method in order to analyze
the digital artefacts left on all accessible devices, such as Mobile phones (e.g., iPhone and Android smartphone) and Desktop systems, running different OS (e.g., Windows and Mac) while using Amazon S3, Google Docs, Dropbox, and Evernote.
Contrary to most of the cloud storage services that are based on open source platforms, Apple users, have their own special cloud storage called \textit{iCloud}. Oestreicher~\cite{oestreicher2014forensically} investigated particularly iCloud service in order to find leftover digital droplets when using native Mac OS X during system synchronization with the cloud. There are also various research studies on several different cloud storage services that we summarized in Table~\ref{table:relatedwork}. We refer the interested reader to~\cite{martini2014cloud, pichan2015cloud} for a comprehensive survey in this regard.


\begin{table}[h]
\caption{A brief overview of the existing cloud storage forensics research studies.}\label{table:relatedwork}
    \begin{center}
    \footnotesize
    \begin{tabular}{@{}lcc@{}}
         \textbf{Cloud Services}&  \textbf{Public Cloud}  & \textbf{Private Cloud}
    \\  \midrule
       \textbf{Dropbox}    & \cite{chung2012digital, federici2014cloud,grispos2013using, martini2015mobile, quick2013dropbox, quick2013forensic, daryabar2016forensic} & \cite{grispos2013using} 
    \\   \textbf{Amazon S3}  &  \cite{chung2012digital}   & 
    \\    \textbf{Evernote}       &   \cite{chung2012digital}  & 
    \\    \textbf{Google Drive}       & \cite{chung2012digital, federici2014cloud, marturana2012case, quick2014google, quick2013forensic, daryabar2016forensic}    &    
    \\   \textbf{SkyDrive}   &  \cite{federici2014cloud, quick2013digital, quick2013forensic}   & 
    \\   \textbf{Box}   &   \cite{grispos2013using, daryabar2016forensic}  &  \cite{grispos2013using}
    \\      \textbf{SugarSync}    & \cite{grispos2013using, shariati2016sugarsync} &  \cite{grispos2013using}   
	\\	\textbf{Amazon Cloud Drive}    & \cite{hale2013amazon} &  
	\\	\textbf{OneDrive}    & \cite{martini2015mobile, daryabar2016forensic} &  
	\\	\textbf{ownCloud}    & \cite{martini2015mobile} & \cite{martini2013cloud}
	\\      \textbf{Flicker}    & \cite{marturana2012case} &     
	\\      \textbf{PicasaWeb}    & \cite{marturana2012case} &     
	\\      \textbf{iCloud}    & \cite{oestreicher2014forensically} &     
	\\      \textbf{UbuntuOne}    & \cite{shariati2015ubuntu}  &     
    \\      \textbf{hubiC}    & \cite{blakeley2015cloud} &     
	\\      \textbf{Mega}    & \cite{daryabar2016mega} &     
	\\      \textbf{Hadoop}    &  &  \cite{cho2012cyber}   
	\\      \textbf{Amazon EC2}    &  &   \cite{dykstra2012acquiring}  
	\\      \textbf{vCloud}    &  &     \cite{martini2014remote}
	\\      \textbf{XtreemFS}    &  & \cite{martini2014distributed}    
	\\      \textbf{Eucalyptus}    &  &  \cite{anwar2011digital}
	\\ \textbf{Amazon AWS}		&\cite{marty2011cloud}	&
\\ \bottomrule
    \end{tabular}
    \end{center}
\end{table}
\textit{\textbf{Outline.}} The rest of the chapter is organized as follows: in Section~\ref{sec:method} we explain the research methodology and experimental setup. Section~\ref{sec:analysis} presents the results of our experimental analysis on pCloud. We answer the question ``What data can be captured from network traffic?" in Section~\ref{sec:traffic}. Finally, Section~\ref{sec:conclusion} concludes the chapter. 

\section{Research Methodology}\label{sec:method}
In order to conduct a reliable digital forensic analysis, we should follow a forensic investigation guideline~\cite{Wilkinson2011, kent2006guide}. In this research study, we performed our forensic investigation based on the framework introduced by Martini and Choo~\cite{martini2012integrated} which is composed of four important stages (Figure~\ref{fig:framework-mart}):   

\begin{figure}[h!t]
	\centering
	\includegraphics[width=0.40\columnwidth]{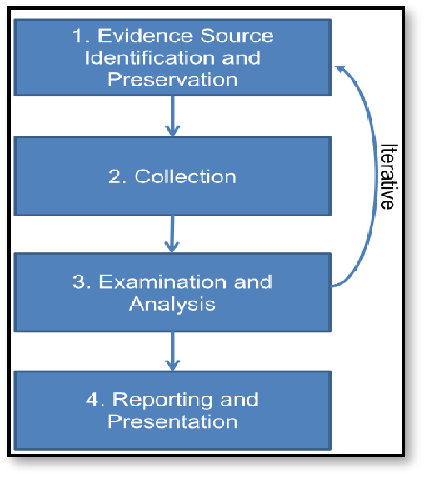}
	\caption{Cloud forensics framework of Martini and Choo~\cite{martini2012integrated}.}
    \label{fig:framework-mart}
\end{figure}

\begin{itemize}

\item[-] \textit{\textbf{Evidence source identification and preservation.}} In this phase, we detect potential sources of evidences. We used \emph{VMware Workstation~10.0.2 build~1744117} to create the Virtual Machines~(VMs) for the experiments. We configured each virtual machine with 1 GB of RAM, and 2 GB hard disk space for Android VM, 15 GB for Ubuntu VM, and 40 GB Windows VM.

\item[-] \textit{\textbf{Collection.}} In this phase, we collected the potential data resources and files in a forensically sound manner.

\item[-] \textit{\textbf{Analysis.}} In this phase, we analyzed the data obtained from the previous phase. We considered keywords such as \textit{``account"},\textit{``password"}, and \textit{``files"} to search for evidence in the memory. This chapter is mainly focused on presenting analysis results of pCloud platform, however, we highlighted the collection and preservation approaches as deemed necessary!

\item[-] \textit{\textbf{Reporting and presentation.}} This phase presents the collected evidences, in such a way that would be acceptable by the court of law. As this chapter is only focused on presenting potential evidences; we just shortly discussed this stage in conclusion.

\end{itemize}

\subsection{Experimental Setup}\label{sec:expSetup}
We conducted our experiments on four different operating systems: a 64~bit Windows~8, Ubuntu~14.04.1~LTS, Android~4.4.2, and iOS~8.1. We considered two different browsers: \textit{Internet Explorer~10.0.9200.16384} and \textit{Google Chrome~39.0.2171.71 m}. We carried out our experiments using the digital forensic research workshop challenge 2013 dataset (DFRWS\footnote{\url{http://www.dfrws.org/2013/challenge/index.shtml}}). We downloaded the dataset on $08^{th}$ December 2014 and evaluated the hash of the dataset to ensure the integrity of the data. The dataset contains a main folder called \texttt{test} including ten directories namely: \texttt{au, b, img, js, ml, msx, pdf, txt, vid, z}. We carried out our investigation taking into account all the files included in all directories.

We utilized \emph{Wireshark~1.12.3} to capture network traffic in all of the platforms and experimental tasks running on them. Furthermore, we used \emph{NetworkMiner~1.6.1} to further analyze the captured network traffic. We captured physical memory in Ubuntu using \emph{memdump~1.01-6-i386}. We used \emph{Hex Workshop~6.7} (6.8.0.5419 /  $1^{st}$ Sep. 2014)  to analyze the captured physical memory of the virtual machines, after the successful execution of each task. One of the main goals of examining this type of application is to determine the possible remnants on different platforms using certain tools, which we explain in the following. 
Apart from the \emph{sqlitebrowser~3.4.0}, we also adopted \emph{iphonebackupbrowser-r38} for Android and iOS. 

\subsubsection{Windows}\label{subsec:wndows}
In order to investigate pCloud remnants on a Windows operating system, we considered two different research directions: i) Windows web browser-based analysis, and ii) Windows application-based analysis.

\begin{figure}[h!t]
    \centering
    \includegraphics[width = .8\columnwidth]{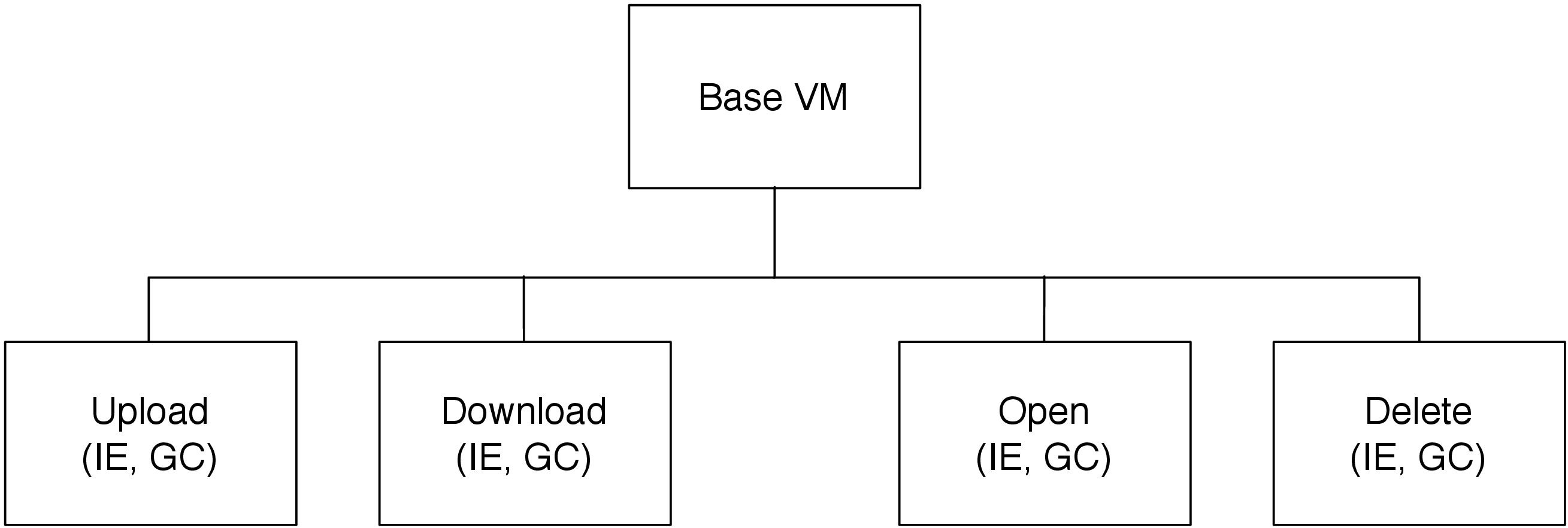}
    \caption{Windows browser-based VMs.}\label{fig:Win-Browser}
\end{figure}

As for the web browser-based investigation, we installed two popular browsers: \emph{Microsoft Internet Explorer~10.0.9200.16384}, and \emph{Google Chrome~39.0.2171.71 m} on four VMs, and performed different tasks specifically to the VM. Figure~\ref{fig:Win-Browser} shows the web browser-based tasks on the Windows VM.
We updated Microsoft Internet Explorer, and installed Google Chrome on the base machine. We then cloned to four other machines for the following tasks: \textit{upload}, \textit{download}, \textit{open} and \textit{delete}. Since it is a browser-based experiment, installation of pCloud was not required, as the experiment will be directly focusing on interacting with the pCloud in the web browser. We used all the folders and files from the DFRWS dataset during each task. For example, we first uploaded all the files during the upload task, and then downloaded back during the download task. Moreover, we captured network traffic during all the tasks.

The main artefacts which are recoverable from web browsers are from their cache and history folders. Therefore, after performing the four aforementioned operations  (i.e., upload, download, open and delete) using the dataset, we analyzed the cache using \textit{NirSoft IECacheview v 1.53} for Internet Explorer and \textit{NirSoft ChromeCacheView v 1.61} for Google Chrome.

\begin{figure}[h!t]
    \centering
    \includegraphics[width = .8\columnwidth]{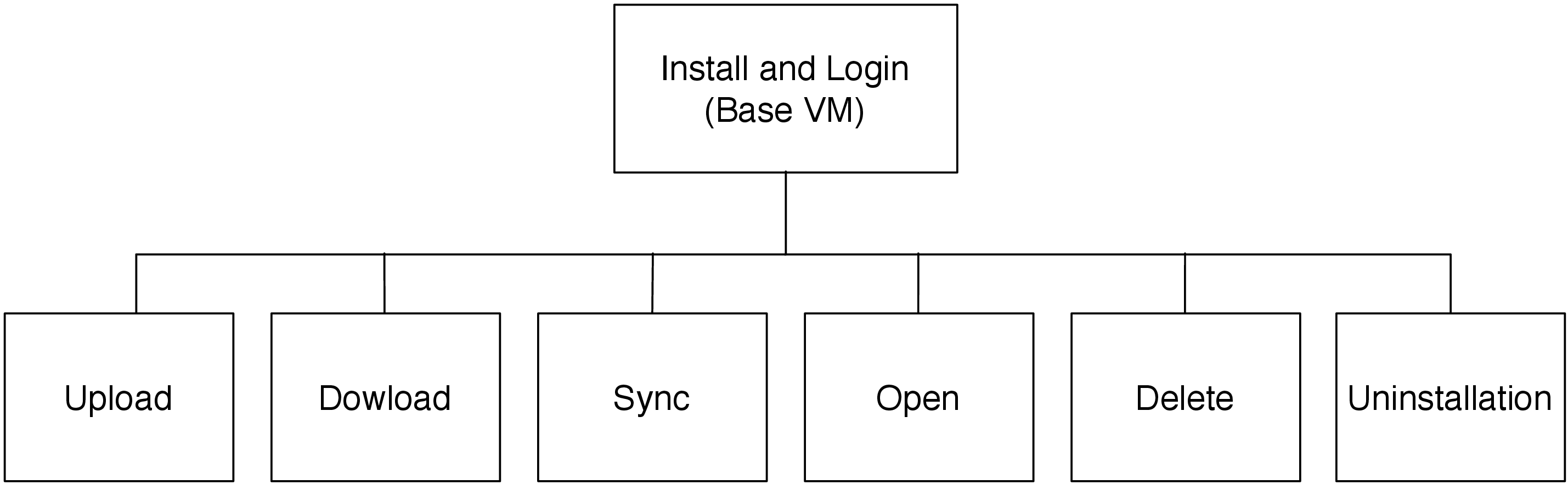}
    \caption{Six different tasks performed on Windows application-based, Android  application-based, iOS-based and Ubuntu-based VMs. It should be noted that, since we performed the same operations for all of the four operating systems, we demonstrated all in one figure. }\label{fig:VM-Apps}
\end{figure}

In order to conduct the windows app-based investigation, we adopted \textit{Windows~8.1 Pro build 9600} with pCloud drive~2.0. We performed six different tasks as it can be seen in Figure~\ref{fig:VM-Apps}.

\subsubsection{Android}\label{subsec:android-app}
In order to access the system folders on Android, the OS needs to be rooted. Without the root access, there is no way of accessing the data which are required to perform the experiments and capturing the internal memory. Also having the root access, we will be able to run certain commands and access system protected files. To interact with the given Android machine, we used a terminal called \textit{Android emulator~1.0.5}. We accessed the system protected files using an application called \textit{Root Browser 
~2.2.3}. With the help of this file browser, we were able to locate different critical artefacts related to the pCloud, such as databases and log files. We used terminal emulator in order to capture the processes which were running in the internal memory (RAM Capture) and also to copy the captured file to the main investigation machine.
We carried out six different experiments on Android based application (Android~4.4.2), which are depicted in Figure~\ref{fig:VM-Apps}. 


\subsubsection{iOS}\label{subsec:iOS}
In order to conduct experiments on iOS, we adopted an iPad mini running \textit{iOS~8.1}. However due to some authentication issues from the owner, we were unable to jail break. We used \textit{iTunes~12.0.1.26}, 64 bit,  to back up the files after performing the tasks which are shown in Figure~\ref{fig:VM-Apps}.
After completion of each task, we took a back up of the whole ipad using \textit{iTunes} with the use of \textit{iPhone Backup Browser~1.2.0.6} (by Google project). We were able to track the changes which was made during the installation procedure of the pCloud.


\subsubsection{Ubuntu}\label{subsec:ubuntu-app}
We adopted the \textit{Ubuntu 14.04.1 LTS} (Trusty Tahr) to carry out our investigation. We installed the pcloud drive~2.0 through the Ubuntu software center. We also performed the uninstallation process through the Ubuntu software center.
We installed the pCloud drive on a main VM which was the ``install and login".
We also cloned this virtual machine for the other tasks: upload, download, sync, open, delete and uninstallation (see Figure~\ref{fig:VM-Apps}). In fact, we cloned these machines in order to avoid the virtual memory being overwritten by the execution of the next task, which would erase the evidence of the previous task with the new evidence of the next task.

The DFRWS dataset contained various types of files, and we used all of them in the experiments. For example, during upload, we uploaded all the folders and files. After the successful execution of each task, we captured the live memory using \textit{memdump~1.01-6}. During
carrying out all the tasks, we also captured the network traffic using \textit{Wireshark~1.12.3}.


\section{Analysis and Findings}\label{sec:analysis}

In this section, we present our experimental findings along with the data anaylsis. In order to analyze the live memory we accessed the VM folders while the VM was powered on. We analyzed this memory, using \textit{Hex Workshop}, after the corresponding task was successfully performed on the VM. 
It should be noted that, if we do not mention a specific action (such as download, upload, sync, or delete), it means there was no evidence which could be used by a forensic investigator for further analysis of that action.

\subsection{Windows Browser-Based Experiments}\label{subsec:windows-web-analysis}
We first set up the base virtual machine in order to conduct the experiments over Windows browser-based. As all the clones had the latest version of the Internet Explorer and Google Chrome, we avoided installing and updating the browsers when conducting experiments. We started the experiments with the ``upload" VM, leading to the ``download" VM, and after that ``open" and ``delete" VMs.

\subsubsection{Upload}
\begin{figure}[h!t]
    \centering
    \includegraphics[width = \columnwidth]{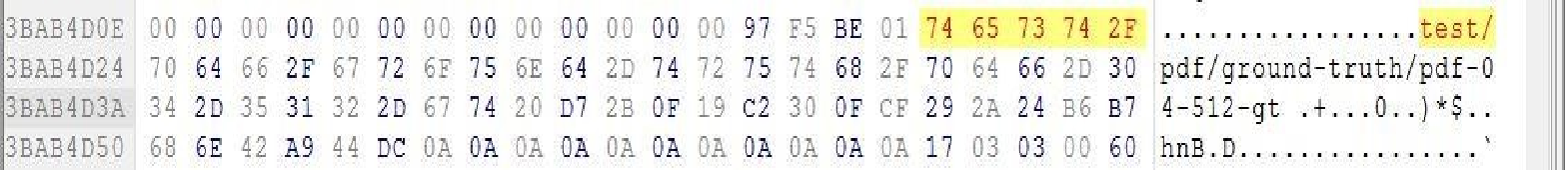}
    \caption{Windows Browser-based -- Uploaded Files}\label{fig:Win-upload}
\end{figure}

We were able to acquire information such as uploaded file names and the user names, which was used to upload the data to the pCloud, using Internet Explorer. As Figure~\ref{fig:Win-upload} shows, we could discover the folder path from the memory.

\subsubsection{Install and Login}
\begin{figure}[h!t]
	\centering
	\begin{subfigure}
	    \centering
		\includegraphics[width=\columnwidth]{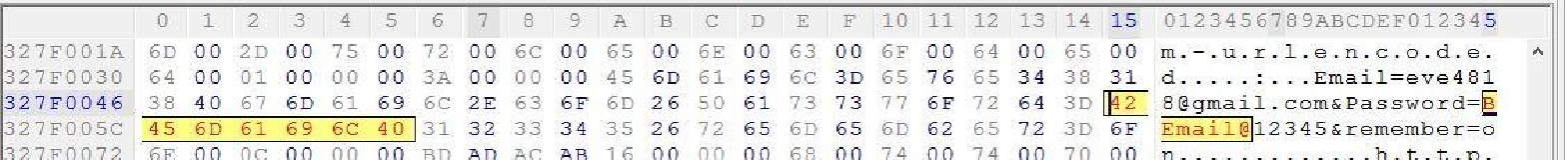}
	\end{subfigure}	
	\hspace{.2cm}
	\begin{subfigure}
	    \centering
		\includegraphics[width=\columnwidth]{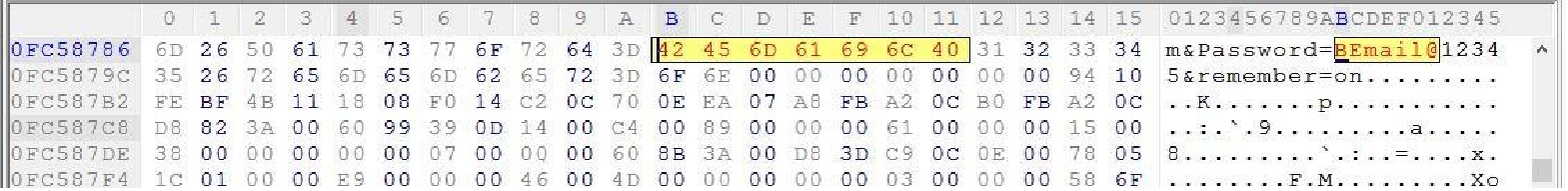}
	\end{subfigure}
	\caption{Windows Browser-based -- Install and Login} 
	\label{fig:Win-insta-logi}
\end{figure}

As it can be seen in Figure~\ref{fig:Win-insta-logi}, the passwords and the email address are clearly discoverable from the physical memory, along with the interested file names and directories. These information are valuable for a forensic examiner. 

We utilized \textit{NirSoft IEPassView~1.32} in order to analyse the saved data files by the Internet Explorer. We found out that Internet Explorer saves the pCloud credentials in the registry. 
However, search results do not reveal any kind of information regarding the credentials in the Internet Explorer cache files, for which we used \textit{IECacheview} to perform analysis. Obtained results indicate that all the uploads went through an encrypted server, making it difficult to gather much information about the uploads.

\begin{figure}[h!t]
    \centering
    \includegraphics[width = \columnwidth]{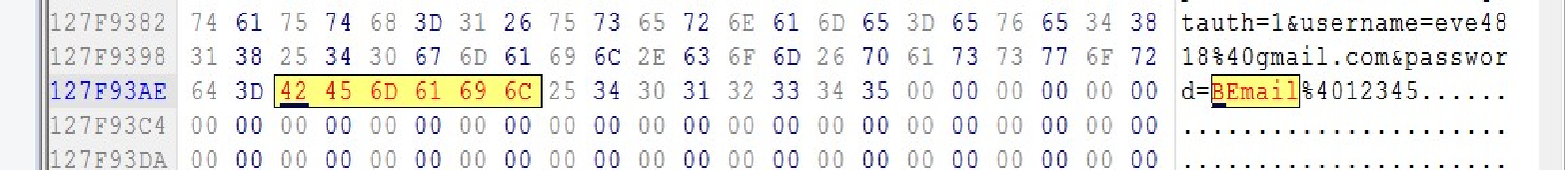}
    \caption{Chrome -- pCloud credentials}\label{fig:Chrome}
\end{figure}

We also analyzed the memory image focusing on the Google Chrome browser. We were able to retrieve remnants such as username and password which was used to access pCloud (Figure~\ref{fig:Chrome}). We could also find evidential data through \textit{ChromeCacheView} for Google Chrome cache, along with the links which were accessed during the tasks.

\subsection{Windows app-Based Experiments}\label{subsec:windows-app-analysis}
In this section we discuss the evidential data we obtained while analysing the pCloud application (app) installed on Windows OS. We explain three different tasks: Install and Login, Delete, and Uninstall.

\subsubsection{Install and Login}
Upon the first installation of the pCloud on Windows, we have traced down the changes that the app made on both file system and the registry of the computer.
The pCloud client created and modified the following address on the disk drive:
\texttt{ Users\textbackslash User\textbackslash Documents\textbackslash pCloud  Sync}.
This address is used to store the pCloud client files, the configuration and some other necessary files.
Other than the system's disk drive, pCloud has created entries in the registry of the Windows.
The Registry entries can be find in the following locations: 
\begin{itemize}

\item \texttt{HKEY\_CURRENT\_USER\textbackslash Software\textbackslash pcloud }

\item \texttt{HKEY\_CURRENT\_USER\textbackslash Software\textbackslash pcloud LTD\textbackslash pCloud Drive} 

\item \texttt{HKEY\_CURRENT\_USER\textbackslash Software\textbackslash Microsoft\textbackslash Windows\textbackslash CurrentVersion\textbackslash Run}

\item \texttt{HKLM\textbackslash SOFTWARE\textbackslash Wow6432Node\textbackslash Microsoft\textbackslash Windows\textbackslash CurrentVersion\textbackslash\\ Uninstall\textbackslash $\lbrace 3e0d7412-ce78-4007-a287-f4a4b42460b2\rbrace$\textbackslash DisplayName: "pCloud Drive"}

\item \texttt{HKLM\textbackslash SOFTWARE\textbackslash Wow6432Node\textbackslash Microsoft\textbackslash Windows\textbackslash CurrentVersion\textbackslash \\ Uninstall\textbackslash
$\lbrace 3e0d7412-ce78-4007-a287-f4a4b42460b2\rbrace$\textbackslash DisplayVersion: "2.0.3.0"}

\item \texttt{HKLM\textbackslash SOFTWARE\textbackslash Wow6432Node\textbackslash Microsoft\textbackslash Windows\textbackslash CurrentVersion\textbackslash \\ Uninstall\textbackslash 
 $\lbrace 3e0d7412-ce78-4007-a287-f4a4b42460b2\rbrace$\textbackslash Publisher: "pCloud~LTD"}

\item \texttt{HKLM\textbackslash SYSTEM\textbackslash CurrentControlSet\textbackslash Services\textbackslash SharedAccess\textbackslash Parameters\textbackslash \\ FirewallPoli cy\textbackslash FirewallRules\textbackslash 
 $\lbrace 9CB654A6-21A1-46DA-A953-0FCB19FE13CA\rbrace$:  "v2.22|Action=Allow|Active=TRUE|Dir=In|\\ Protocol=6|App=\% ProgramFiles(x86)\% pCloud Drive\textbackslash pCloud.exe| \\ Name=pCloud|Edge=TRUE|"}

\item \texttt{HKLM\textbackslash SOFTWARE\textbackslash Microsoft\textbackslash Windows\textbackslash CurrentVersion\textbackslash Installer\textbackslash UserData\textbackslash \\ S-1-5-18\textbackslash Products\textbackslash 2CB735048C972D445A5864132F3A0314\textbackslash InstallProperties\textbackslash \\ DisplayName: "pCloud Drive"}

\item \texttt{HKLM\textbackslash SOFTWARE\textbackslash Microsoft\textbackslash Windows\textbackslash CurrentVersion\textbackslash Installer\textbackslash UserData\textbackslash \\ S-1-5-18\textbackslash Products\textbackslash 2CB735048C972D445A5864132F3A0314\textbackslash InstallProperties\textbackslash \\ Publisher:
"pCloud LTD"}

\item \texttt{HKLM\textbackslash SOFTWARE\textbackslash Wow6432Node\textbackslash Microsoft\textbackslash Windows\textbackslash CurrentVersion\textbackslash Uninstall\textbackslash 
\\ $\lbrace 3e0d7412-ce78-4007-a287-f4a4b42460b2\rbrace$ \textbackslash QuietUninstallString: "\%ProgramData\% PackageCache\textbackslash $\lbrace 3e0d7412-ce78-4007-a287- f4a4b42460b2\rbrace$ \textbackslash pCloud Drive.exe /uninstall /quiet"}

\item \texttt{HKLM\textbackslash SOFTWARE\textbackslash Microsoft\textbackslash Windows\textbackslash CurrentVersion\textbackslash Installer\textbackslash UserData\textbackslash \\ S-1-5-18\textbackslash Components\textbackslash B8991F4234EFEBC4F8A2180B2B003A2C\textbackslash \\ 2CB735048C972D445A5864132F3A0314: "01:\textbackslash Software\textbackslash pCloud\textbackslash AppPath"}

\item \texttt{HKLM\textbackslash SOFTWARE\textbackslash Classes\textbackslash CLSID\textbackslash $\lbrace 0b73fac-351f-3948-9d8a-1dad9d870193\rbrace$ \textbackslash InprocServer32\textbackslash CodeBase:file:///\%ProgramFiles(x86)\%pCloudDrive/\\ ContextMenuHandler.DLL}

\end{itemize}

\begin{figure}[h!t]
    \centering
    \includegraphics[width = \columnwidth]{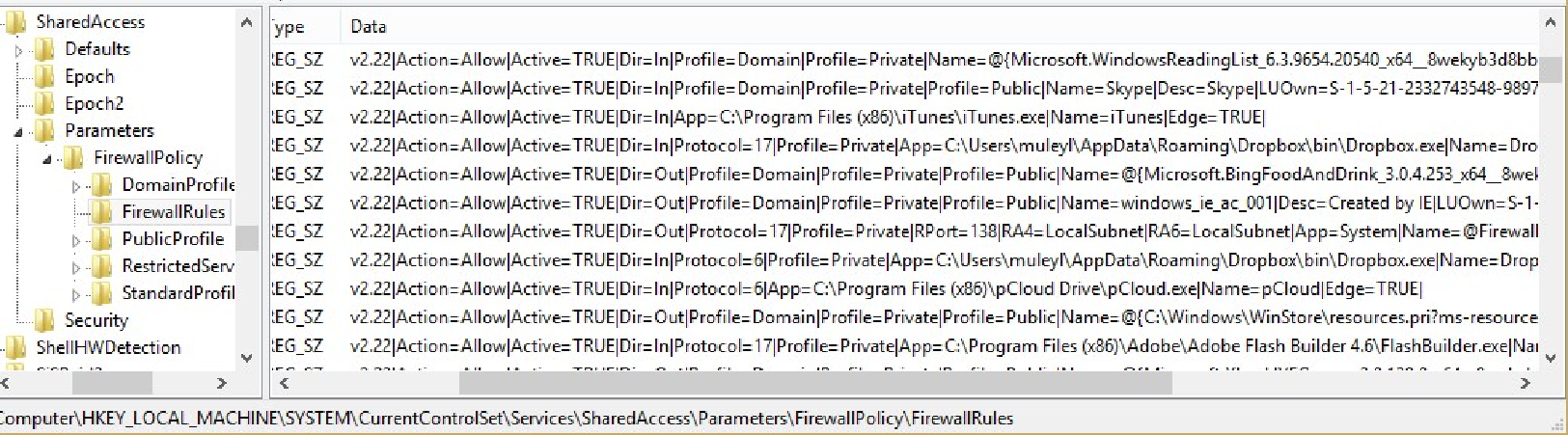}
    \caption{Windows app-based registry changes in firewall folder}\label{fig:win-app-registry}
\end{figure}

We found out that pCloud creates some files in the \textit{Run} and also \textit{Uninstall} folders of the registry. 
Other than changes in the registry and local hard drive, we noticed changes in the rules for Windows Firewall in order to solve the issues that may happen while connecting to the pCloud Servers (Figure~\ref{fig:win-app-registry}).

After reviewing the memory dump images from the Windows machine, which pCloud client was installed on, we found out that we are unable to find any sort of plain text passwords. However, we have successfully found usernames within the memory dump. 

\subsubsection{Delete}

\begin{figure}[h!t]
    \centering
    \includegraphics[width = .8\columnwidth]{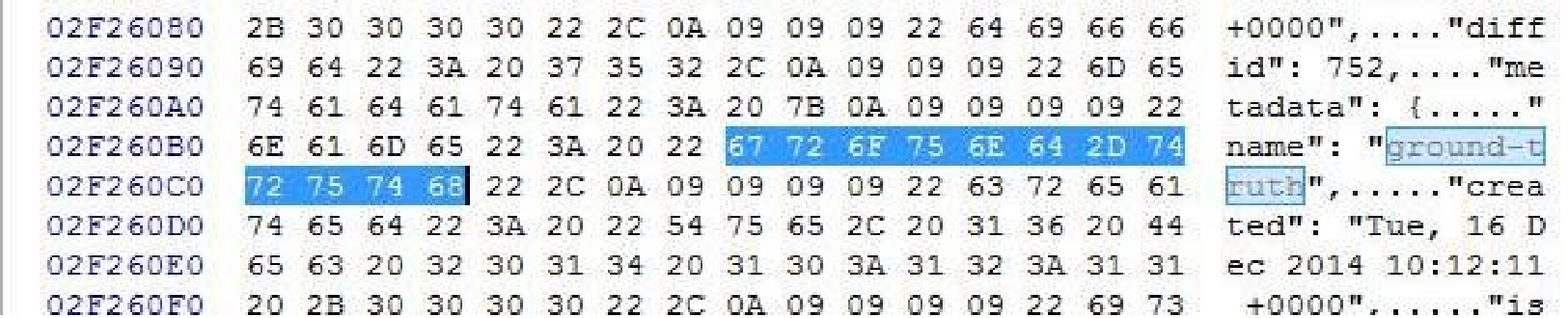}
    \caption{Windows app-based deleted files}\label{fig:win-app-delete}
\end{figure}

In order to analyse the effect of the ``Delete" action, we deleted some files. We recognized that it is still possible to find some traces of the deleted file names within the memory dump (Figure~\ref{fig:win-app-delete}).

\subsubsection{Uninstall}

\begin{figure}[h!t]
    \centering
    \includegraphics[width =.8 \columnwidth]{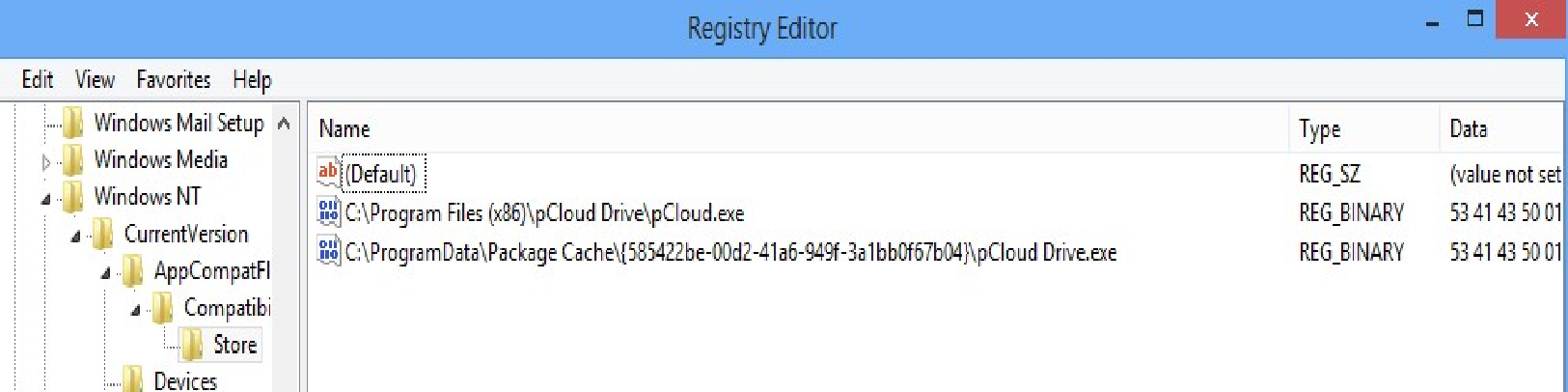}
    \caption{Windows app-based uninstall registry change}\label{fig:win-app-uninstall}
\end{figure}

After the uninstallation process of the pCloud from the VM, we detected two registry entries (Figure~\ref{fig:win-app-uninstall}).

Other than the changes in registry, there were some files left on the disk after uninstallation of the application, which were located at: \texttt{\textbackslash User\textbackslash AppData\textbackslash local\textbackslash \\ pCloud}. Moreover, we found out that pCloud client stores every information such as ``account information" and the ``files summery" in a database called \textit{Data.db} on the computer. This database uses \textit{sqlite dbms }system. From this database file, we were able to extract different kinds of data such as ``uploaded file names", and ``usernames" which the client used to access the pCloud. Moreover, by analysing the database, we found a table called ``file" which keeps all the stored files names. We could find all the files, which we created on our pCloud account. Furthermore, we were able to recover our pCloud account information, such as ``userid" and ``username" in a table called ``settings".

\subsection{Android app-Based Experiments}\label{subsec:android-app-analysis}
In this section, we provide our experimental results related to the pCloud application when using Android OS. We considered three tasks: Install and Login, Upload and Uninstall. 

\subsubsection{ Install and Login}

Once the pCloud was installed on the Android platform, the following two folders were created:

\begin{itemize}
\item \texttt{ /Device/data/data/com.pcloud.pcloud}
\item \texttt{/emulated/0/.pcloud/}
\end{itemize}

By using ``Root Browser", it is possible to locate those folders after completion of the installation process. Moreover, an examination of the memory capture revealed useful information other than user login details, such as folder paths, its database location and other pCloud related
information. We recognized that the database for pCloud was stored in the following locations:

\begin{itemize}
\item \texttt{/data/data/com.pcloud.pcloud/databases/PCloudDB/}
\item  \texttt{/data/data/com.pcloud.pcloud/databases/PCloudDB-journal/}
\end{itemize}

\begin{figure}[h!t]
    \centering
    \includegraphics[width = .9\columnwidth]{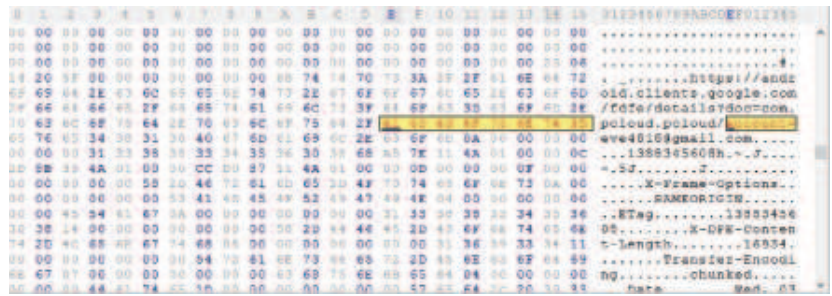}
    \caption{Android -- account details}\label{fig:android-acc}
\end{figure}

Analysing the database using \textit{Sqlitebrowser~3.4.0}, it is possible to find ``usernames", ``email quota", and ``tables", which are related to pCloud communications. Once the pCloud was installed, We logged in from the account which we created previously. Then, the system analyser dumped the whole memory of Android and sent it to analysis machine for further analysis.
We analyzed the memory using \textit{Hex Workshop~6.7}. In order to find data related to the user account, we used a search string (i.e., ``account="). This way, we could identify the account which we had registered for the cloud storage (Figure~\ref{fig:android-acc}). 

\begin{figure}[h!t]
    \centering
    \includegraphics[width = 0.9\columnwidth]{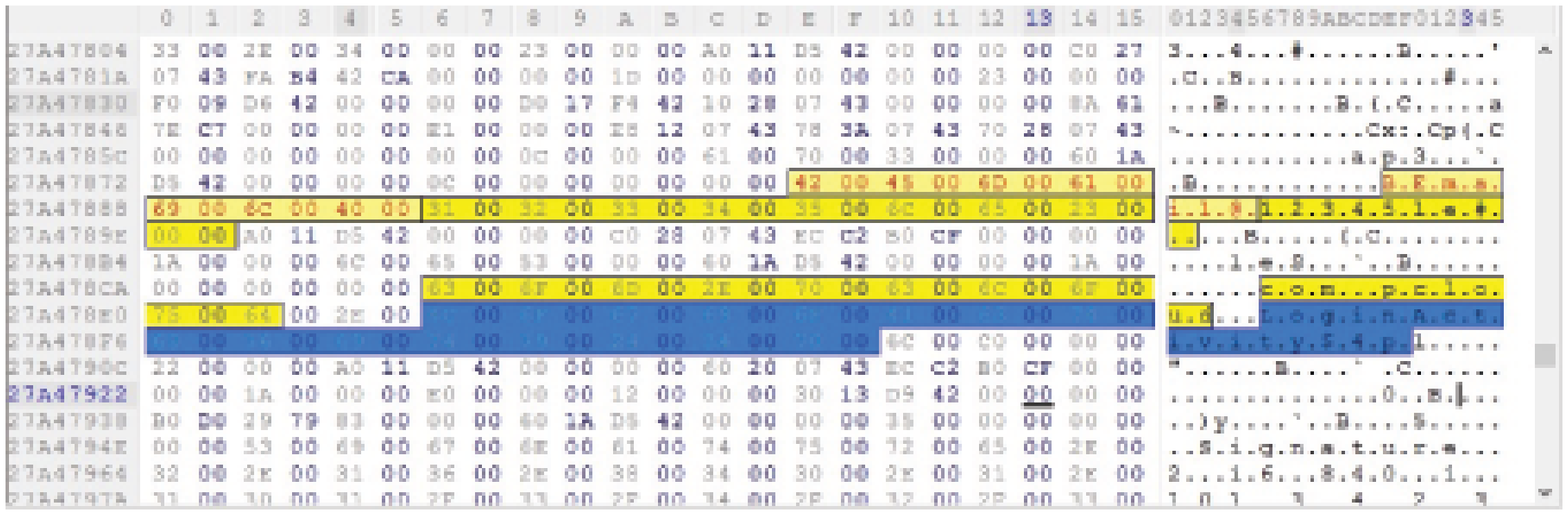}
    \caption{Android -- pCloud suspected credentials (highlighted part)}\label{fig:android-pcloud}
\end{figure}

Upon finding the registered account, we used it in order to check if it is possible to find more credentials' details! Figure~\ref{fig:android-pcloud} shows the extracted artefacts highlighted in yellow.

\subsubsection{Upload}

\begin{figure}[h!t]
    \centering
    \includegraphics[width =0.9 \columnwidth]{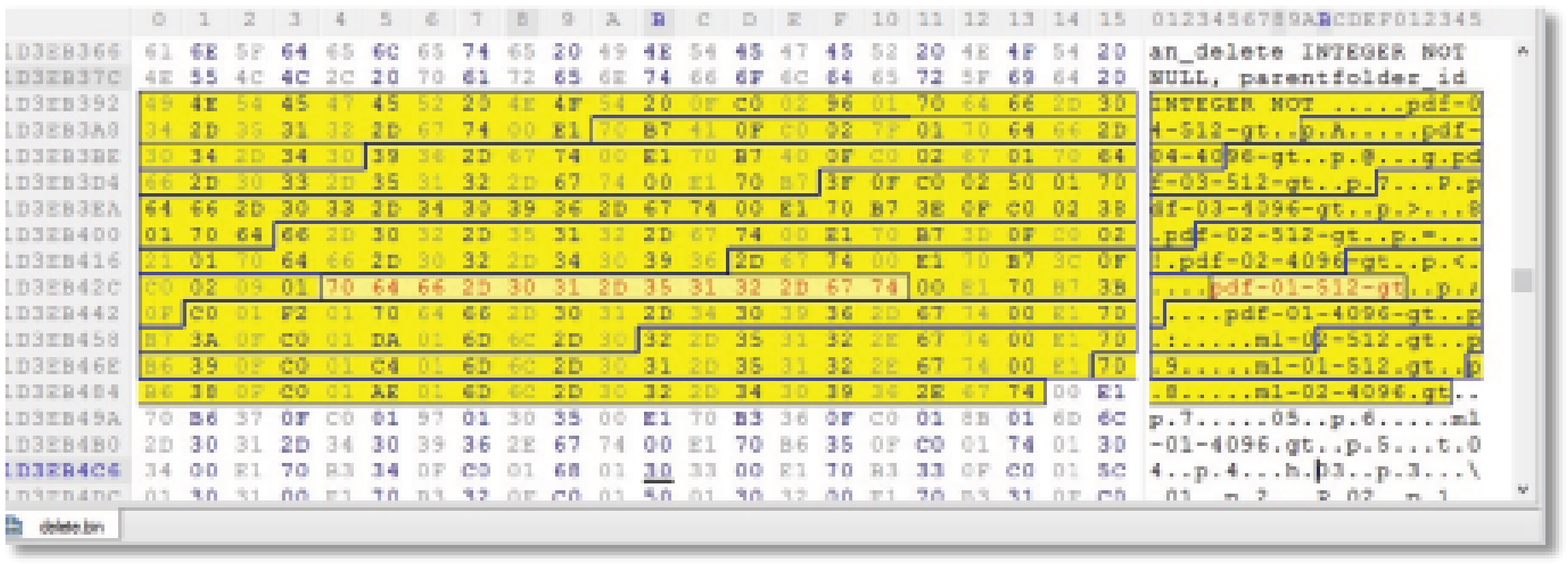}
    \caption{Android based -- uploaded files}\label{fig:android-upload}
\end{figure}

Considering the upload task, we could recover some of the files, which were uploaded to the pCloud, from the memory capture. To this end, we used the search string ``file". A part of the files are demonstrated in Figure~\ref{fig:android-upload} (the highlighted parts). 

\subsubsection{Uninstall}

In order to investigate the possible evidential data which could be remained on the memory after uninstallation of the application, we uninstalled the pCloud application and captured the memory. We could recover some of the folders which were already created in the installation process. We were also able to recover some of the details by accessing the default browser in Android. We logged into the pCloud service using Android default web browser, then we analyzed the cache file, and browser history. We could recover evidences such as website information, and some cookie files regarding the access of pCloud.

\subsection{iOS Based Experiments}\label{subsec:ios-analysis}

Examining iOS for finding possible pCloud artefacts was difficult due to the complexity of the OS, compared to other operating systems. Moreover, we were unable to Jail break iOS. Therefore, we adopted backup investigation method to detect the exact location of the installed pCloud.

\begin{figure}[h!t]
    \centering
    \includegraphics[width = .9\columnwidth]{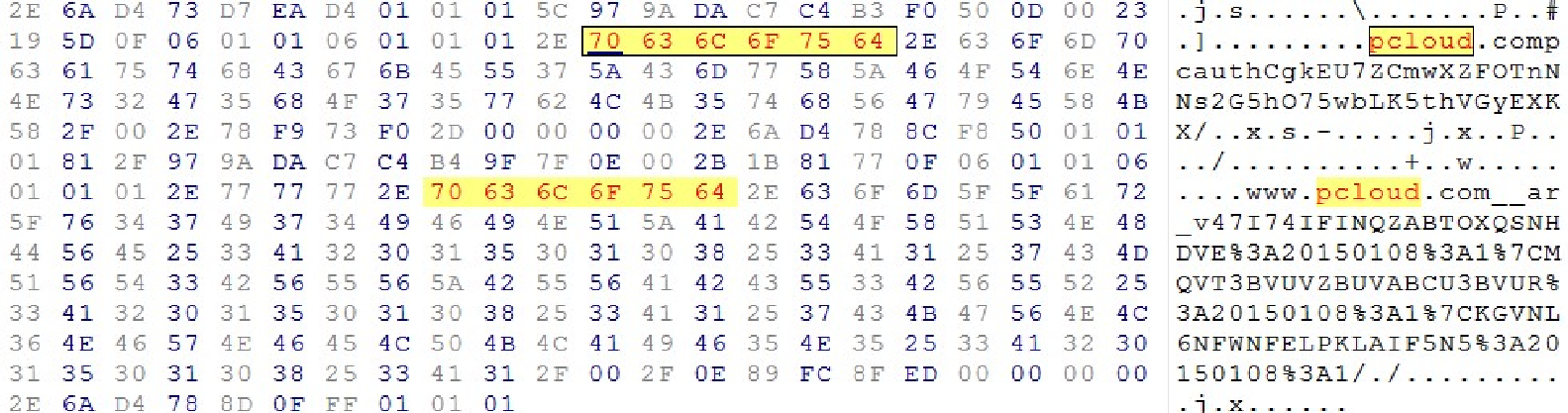}
    \caption{Android Based -- cookies and web page info}\label{fig:android-cookie}
\end{figure} 

\begin{figure}[h!t]
    \centering
    \includegraphics[width = .7\columnwidth]{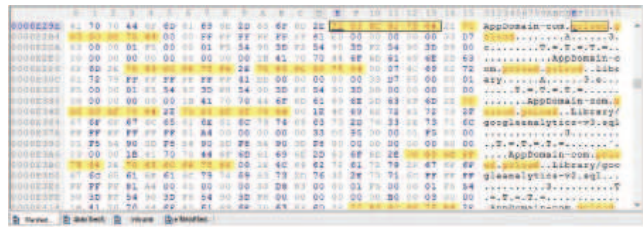}
    \caption{iOS pcloud folder paths}\label{fig:ios-pcloud}
\end{figure}

 Upon installation of the pCloud on iOS, the folders/files which are depicted in Figure~\ref{fig:ios-pcloud} were created in the following locations:

\begin{itemize}

\item \texttt{Library/Preferences/com.pcloud.pcloud.plist}

\item \texttt{Library/googleanalytics-v2.sql}

\item \texttt{Library/googleanalytics-v3.sql}

\item \texttt{Library/Application Support/p.db} 
\end{itemize}

\begin{figure}[h!t]
    \centering
    \includegraphics[width = .9\columnwidth]{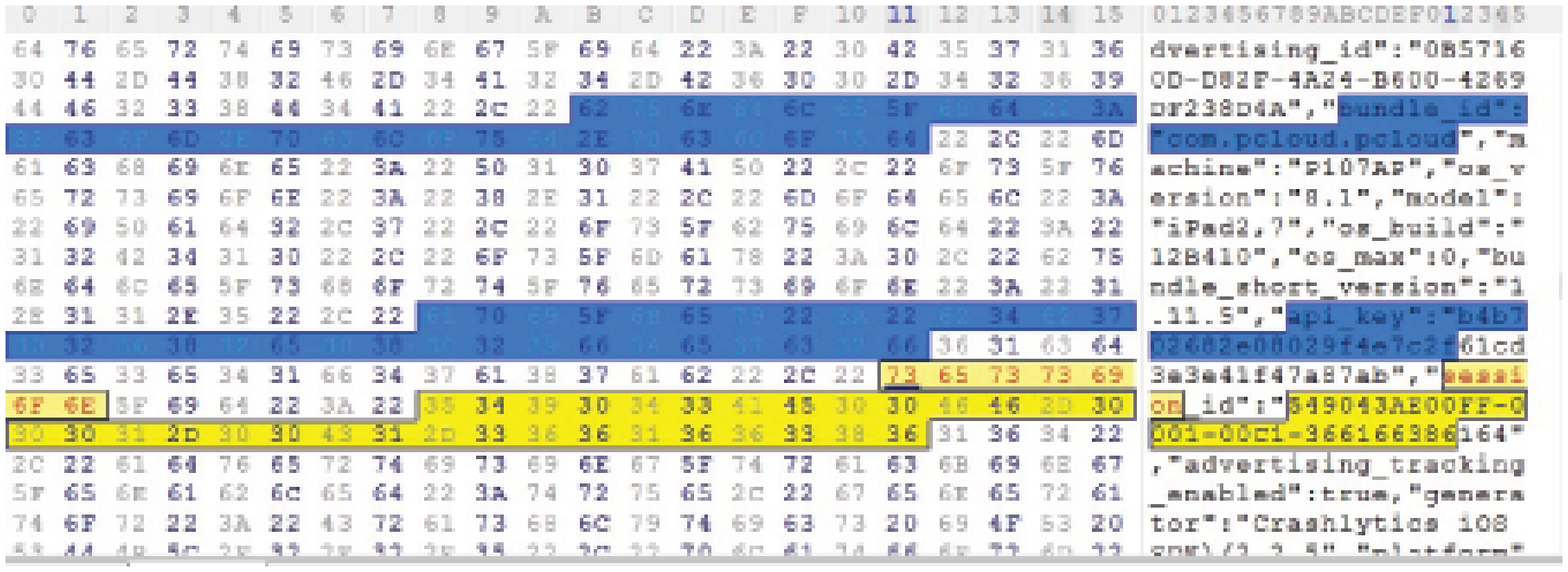}
    \caption{iOS based -- API key, Session ID Findings}\label{fig:ios-API}
\end{figure}

During the analysis process of the iOS backup files, we didn't find any login details related to pCloud. However, we obtained some information such as ``session ID" (type of cookie which the web servers store for a specific user for a duration of time), and ``API key" (a code passed to the computer to identify the calling program to its user), which then could be useful for Forensic Investigations (see Figure~\ref{fig:ios-API}). Furthermore, we could obtain information such as pCloud installation directory location.

\subsubsection{Upload and Uninstallation}

\begin{figure}[h!t]
	\centering
	\begin{subfigure}
	    \centering
		\includegraphics[width=.7\columnwidth]{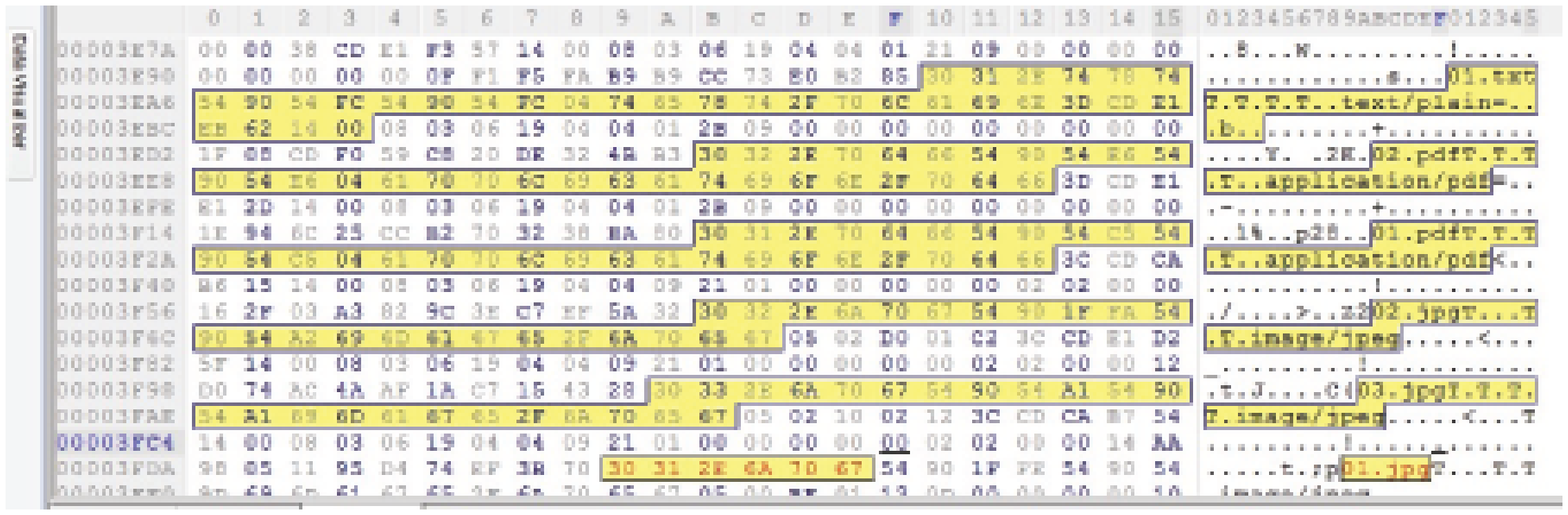}
	\end{subfigure}	
	\hspace{.5cm}
	\begin{subfigure}
	    \centering
		\includegraphics[width=.7\columnwidth]{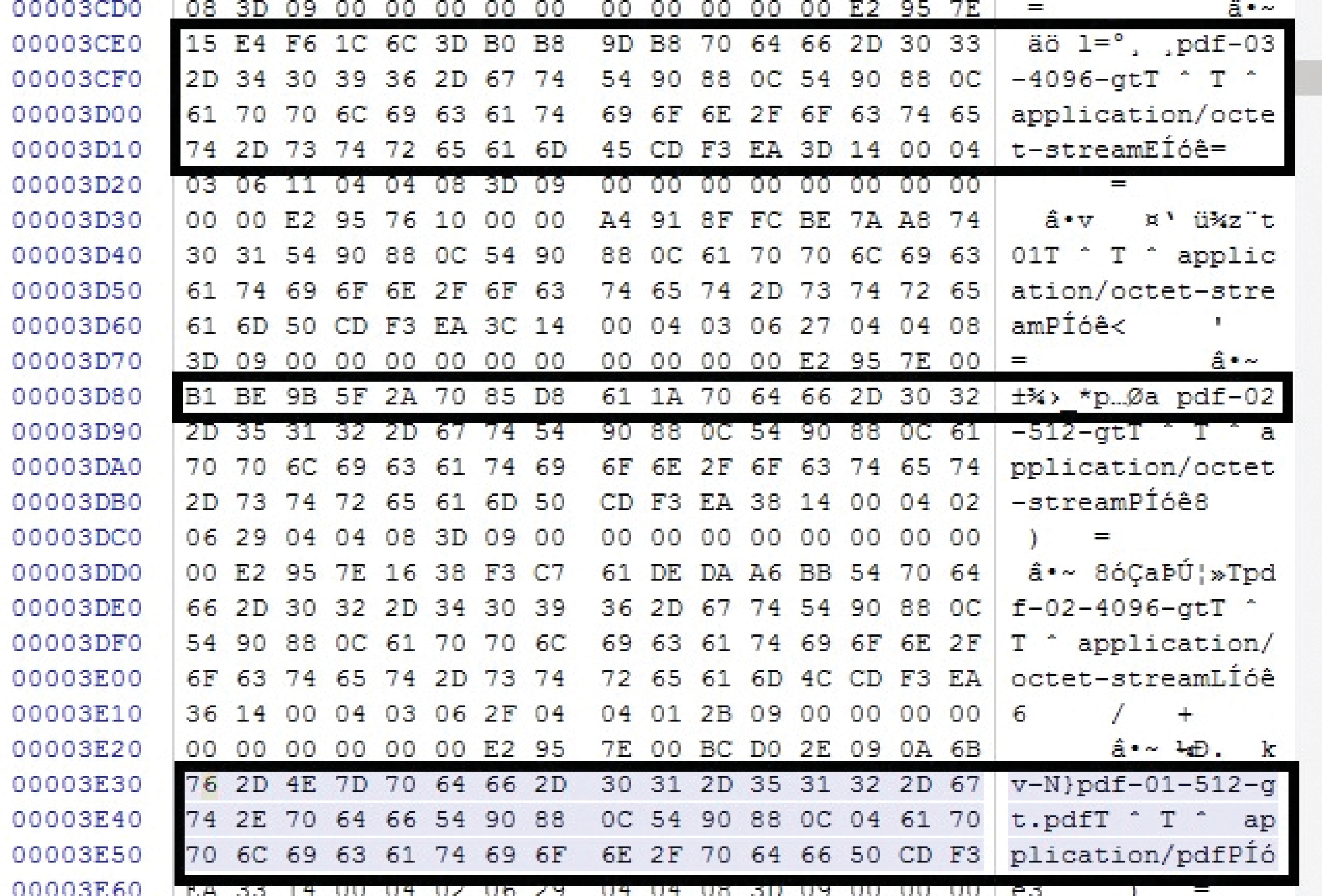}
	\end{subfigure}
	\caption{iOS based uploaded file names found in the backup files} 
	\label{fig:ios-upload}
\end{figure}

Even though we did not obtain pCloud login details on iOS, we could  detect some useful information such as ``uploaded files names" (as highlighted in Figure~\ref{fig:ios-upload}). Moreover, upon uninstallation we could recover some of the deleted files.
In order to access such information, we used several search strings such as common file types, for instance ``\texttt{.jpg}" and ``\texttt{.pdf}".

\subsection{Ubuntu app-Based Experiments}\label{subsec:ubuntu-app-analysis}

During the experimental study on Ubuntu, we installed the pCloud drive~2.0 on base VM, and logged in. Then, we cloned it for several tasks, which we carried out in the following sequential manner: upload, download, sync, open, delete and uninstall.
We analyzed all the acquired memory dump files using \textit{hex workshop}.  We found quite a number of evidences in the memory. These evidences are clearly useful for digital investigators in order to get to know the ``username", ``password", and ``files names" of the victim or suspects.

\begin{figure}[h!t]
    \centering
    \includegraphics[width = .8\columnwidth]{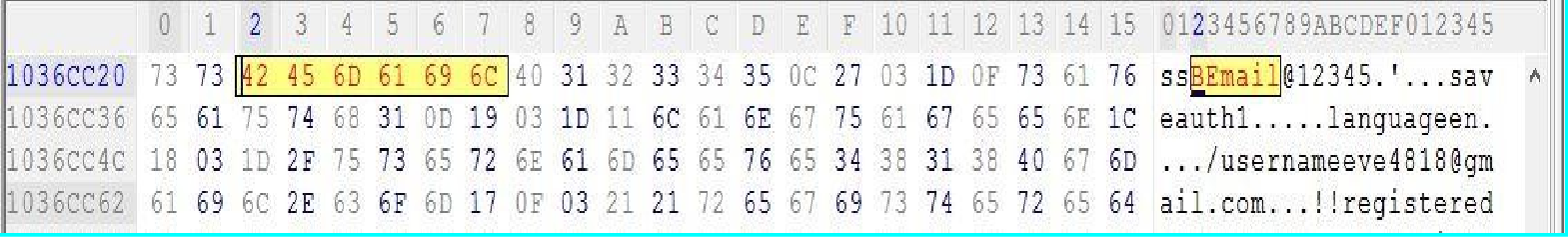}
    \caption{Ubuntu -- revealed credentials during install and login tasks. }\label{fig:Ubuntu-install}
\end{figure}

\begin{figure}[h!t]
    \centering
    \includegraphics[width = .8\columnwidth]{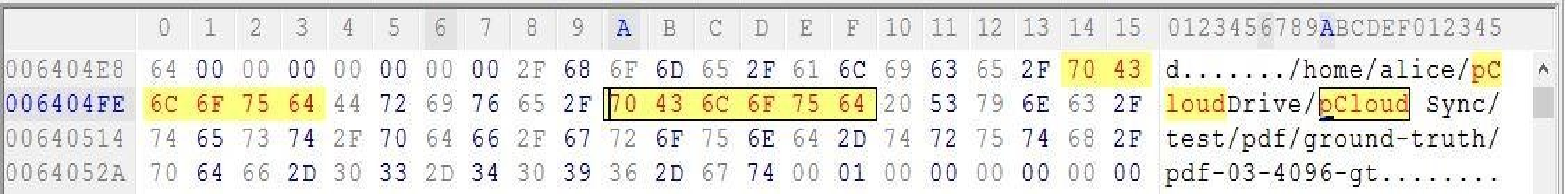}
    \caption{Ubuntu -- extracted information during upload process.}\label{fig:Ubuntu-upload}
\end{figure}

\begin{figure}[h!t]
    \centering
    \includegraphics[width = .8\columnwidth]{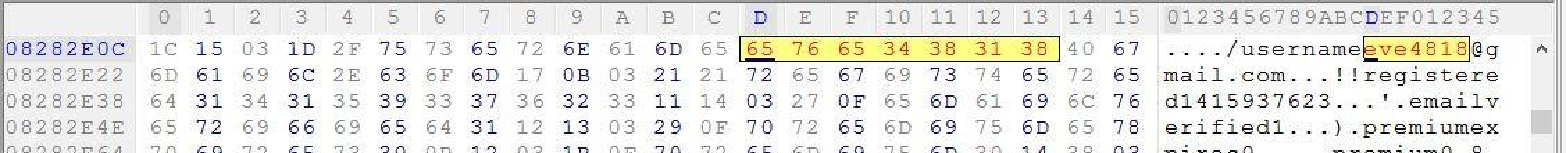}
    \caption{Ubuntu -- recovered username after deletion task. }\label{fig:Ubuntu-delete}
\end{figure}

As it is demonstrated in Figure~\ref{fig:Ubuntu-install}, it is possible to recover the ``username" and ``password" of the user during installation and login process. These information have high forensic value to the forensic examiners as it shows the credentials of the victims/suspects. Moreover, as Figure~\ref{fig:Ubuntu-upload} shows, we can retrieve the uploaded file names and the file path from the memory dump. We could also retrieve the same evidences as the ones extracted from the ``upload” VM memory dump, during the sync and download tasks. 
As it is depicted in Figure~\ref{fig:Ubuntu-delete}, after the deletion of the files from the app, it is possible to recover ``username" 
from the memory dump. This evidence can also help the forensic examiners to identify the credentials that were used.

\section{Network Traffic}\label{sec:traffic}

Compared to the evidential data recovered from the storage and memory, we could obtain relatively limited amount of data by analysing the network traffic.
This is mostly because pCloud uses encrypted connections, such as \textit{TLSv1.2} and \textit{HTTPS} over \textit{SSL} certificates, which are then provided by external vendors. During the download and upload tasks, an encrypted connection is established with protocol HTTPS.

In Table~\ref{table:IP-Address}, we show some relevant IP addresses to pCloud, which we could recover during the Internet Explorer experiment. We can conclude that all the connections to these hosts were over TCP port~443, and used a \textit{TLSv1.2} Encryption.
Apart from these IP addresses, we were able to track the service providers for SSL certificates, along with the main login IP address/URL which we could use as forensic investigators for further analysis.
The recovered SSL Certifcate providers list is as follows:

\begin{itemize}
\item \texttt{http://silver-server-g2.ocsp.swisssign.net/ \\ D3446FD9FE7AFCDEAC1C7AA2210D64FA65B0D782}
\item \texttt{http://crl.swisssign.net/D3446FD9FE7AFCDEAC1C7AA2210D64FA65B0D782}
\item\texttt{ldap://directory.swisssign.net/ \\ CN=D3446FD9FE7AFCDEAC1C7AA2210D64FA65B0D782\% \\ 2CO=SwissSign2CC=CH?certificateRevocationList?base?objectClass=cR LDistributionPoint}
\end{itemize}

\section{Conclusion}\label{sec:conclusion}
In this chapter, by analysing pCloud as a case study we demonstrated the possibility to recover a numerous amount of residual evidences from this platform. We analyzed the pCloud on several operating systems (i.e., Windows, Android, iOS, Ubuntu) considering different tasks (such as, install, login, upload, download, uninstall). We showed that all the pCloud credentials could be extracted along with the files that were used for storage. Even though the network connections were encrypted, some of the credentials used in almost all platforms were in plain text format which is an added advantage for forensic investigators. However we were only able to collect login credentials by capturing the live memory at the time of installation of the pCloud service. So it is highly recommended for forensic investigators to capture the memory at the time of installation. 

Our presented research study in this chapter may pave the way for forensics examiners investigating pCloud and other cloud storage platforms. In future, researchers can use similar investigation method to retrieve other cloud platforms remnants. Extending presented approach for detecting evidences of different platforms over cloud, such as investigating mobile devices connected to the cloud~\cite{mohtasebi2011smartphone, parvez2011framework}, investigation of cloud-based social networking platforms~\cite{mohtasebi2011defusing, norouzizadeh2015investigating}, and cloud malware forensics~\cite{daryabar2011malware, daryabar2011investigation, shaerpour2013trends, dezfouli2013survey, damshenas2013survey} would be interesting future works. Moreover, analyzing legal and privacy implications of conducting cloud forensics~\cite{daryabar2013survey, dehghantanha2014privacy} and developing relevant solutions could further opportunities for real-world utilization of cloud investigation techniques.

 \begin{table}[h]
\caption{Recovered IP addresses during the Internet Explorer experiments.}\label{table:IP-Address}
\centering
 	\footnotesize
 	\begin{tabular}{| l |  p{.4\textwidth} | l| }
 		\hline 
 		 \textbf{IP Address} & \textbf{Host Name} &  \textbf{Activity} \\
 		\hline \hline
 		$74.120.8.17/25/18/24/23/26$ & \texttt{binapi.pcloud.com} & Install and Login \\ \hline
        $74.120.8.24/25/17/18/23/26$ & \texttt{binapi.pcloud.com} & \multirow{ 2}{*}{Uninstall} \\
        $74.120.8.56$ & \texttt{C47.pcloud.com} &  \\ \hline
        $74.120.8.26/25/24/18/23/17$ & \texttt{binapi.pcloud.com} & \multirow{ 11}{*}{Upload}\\
$74.120.8.28$ & \texttt{C1.pcloud.com} & \\
$74.120.8.41$ & \texttt{C19.pcloud.com} &  \\
$74.120.8.56$ & \texttt{C47.pcloud.com} &\\
$74.120.8.64$ & \texttt{C54.pcloud.com} &\\
$74.120.8.73$ & \texttt{C61.pcloud.com} &\\
$74.120.8.89$ & \texttt{C72.pcloud.com} &\\
$74.120.8.92$ & \texttt{C75.pcloud.com} &\\
$74.120.8.96$ & \texttt{C79.pcloud.com} &\\
$74.120.8.100$ & \texttt{C82.pcloud.com} &\\
$74.120.8.133$ & \texttt{C94.pcloud.comb} & \\ \hline
$74.120.8.77$ & \texttt{a2.pcloud.com},
\texttt{translate.pcloud.com} & \multirow{ 3}{*}{Upload}\\
$74.120.8.15/7/6/12/13$ & \texttt{api.pcloud.com},
\texttt{my.pcloud.com} & \\
$74.120.8.14$ & \texttt{api.pcloud.com},
\texttt{my.pcloud.com}, \texttt{api8.pcloud.com} &  \\ \hline
$74.120.8.77$ & \texttt{a2.pcloud.com}, \texttt{translate.pcloud.com} & \multirow{ 4}{*}{Open} \\ 
$74.120.8.15/7/6/12/13$ & \texttt{api.pcloud.com}, \texttt{my.pcloud.com} &\\
$74.120.8.14$&
\texttt{api.pcloud.com}, \texttt{my.pcloud.com}, \texttt{api8.pcloud.com} &\\
$74.120.8.34$&
\texttt{c15.pcloud.com} &\\
 		\hline
        $74.120.8.15/7/6/12/13$&
\texttt{api.pcloud.com}, \texttt{my.pcloud.com} &\multirow{ 3}{*}{Delete}\\
$74.120.8.14$ &
\texttt{api.pcloud.com},
\texttt{my.pcloud.com}, \texttt{api8.pcloud.com} &\\
$74.120.8.34$ &
\texttt{c15.pcloud.com} &\\ \hline
 	\end{tabular}
 \end{table}

\newpage
\section*{References}

\bibliography{mybibfile}

\end{document}